\def\papertitle{Network Bending of Diffusion Models for Audio-Visual Generation}
\def\paperauthorA{Luke Dzwonczyk}
\def\paperauthorB{Carmine Emanuele Cella}
\def\paperauthorC{David Ban}

\documentclass[twoside,a4paper]{article}
\usepackage{etoolbox}
\usepackage{dafx_24}
\usepackage{amsmath,amssymb,amsfonts,amsthm}
\usepackage{euscript}
\usepackage[T1]{fontenc}
\usepackage[utf8]{inputenc}
\usepackage{ifpdf}
\usepackage[english]{babel}
\usepackage{caption}
\usepackage{subcaption}
\usepackage{color}

\input glyphtounicode
\ninept

\newcounter{numauth}
\newcounter{listcnt}
\newcommand\authcnt[1]{\ifdefined#1 \stepcounter{numauth} \fi}

\newcommand\addauth[1]{
\ifdefined#1 
\stepcounter{listcnt}
\ifnum \value{listcnt}<\value{numauth}
\appto\authorslist{, #1}
\else
\appto\authorslist{~and~#1}
\fi
\fi}
\authcnt{\paperauthorB}
\authcnt{\paperauthorC}
\authcnt{\paperauthorD}
\authcnt{\paperauthorE}
\authcnt{\paperauthorF}
\authcnt{\paperauthorG}
\authcnt{\paperauthorH}
\authcnt{\paperauthorI}
\authcnt{\paperauthorJ}
\def\authorslist{\paperauthorA}
\addauth{\paperauthorB}
\addauth{\paperauthorC}
\addauth{\paperauthorD}
\addauth{\paperauthorE}
\addauth{\paperauthorF}
\addauth{\paperauthorG}
\addauth{\paperauthorH}
\addauth{\paperauthorI}
\addauth{\paperauthorJ}

\usepackage{times}

\newif\ifpdf
\ifx\pdfoutput\relax
\else
   \ifcase\pdfoutput
      \pdffalse
   \else
      \pdftrue
\fi

\ifpdf 
  \usepackage[pdftex,
    pdftitle={\papertitle},
    pdfauthor={\authorslist},
    pdfsubject={Proceedings of the 27th International Conference on Digital Audio Effects (DAFx24)},
    colorlinks=false, 
    bookmarksnumbered, 
    pdfstartview=XYZ 
  ]{hyperref}
  \usepackage[pdftex]{graphicx}
\else 
  \usepackage[dvips]{epsfig,graphicx}
  \usepackage[dvips,
    pdftitle={\papertitle},
    pdfauthor={\authorslist},
    pdfsubject={Proceedings of the 27th International Conference on Digital Audio Effects (DAFx24)},
    colorlinks=false, 
    bookmarksnumbered, 
    pdfstartview=XYZ 
  ]{hyperref}
\fi
\usepackage[hypcap=true]{caption}
\title{\papertitle}

\affiliation{
\paperauthorA\,\sthanks{Thanks to the predecessors for the templates}}
{\href{https://dafx24.surrey.ac.uk}{Institute of Sound Recording} \\ University of Surrey\\ Guildford, UK\\
{\tt \href{mailto:dafx24@surrey.ac.uk}{dafx24@surrey.ac.uk}}
}
\affiliation{
\paperauthorA\,\sthanks{Thanks to the predecessors for the templates}and \paperauthorB \,\sthanks{This work was supported by the XYZ Foundation}}
{\href{https://dafx24.surrey.ac.uk}{Institute of Sound Recording} \\ University of Surrey\\ Guildford, UK\\
{\tt \href{mailto:dafx24@surrey.ac.uk}{dafx24@surrey.ac.uk}}
}
\affiliation{
\paperauthorA\ and \paperauthorB\ and \paperauthorC,\thanks{\vspace{-3mm}}}
{Center for New Music and Audio Technologies \\ University California, Berkeley \\ Berkeley, CA, USA\\
{\tt \href{mailto:dz.luke@berkeley.edu}{dz.luke@berkeley.edu}}
}
\begin{document}
\ifpdf 
  \DeclareGraphicsExtensions{.png,.jpg,.pdf}
\else  
  \DeclareGraphicsExtensions{.eps}
\fi


\maketitle

\begin{abstract}
In this paper we present the first steps towards the creation of a tool which enables artists to create music visualizations using pre-trained, generative, machine learning models. First, we investigate the application of network bending, the process of applying transforms within the layers of a generative network, to image generation diffusion models by utilizing a range of point-wise, tensor-wise, and morphological operators. We identify a number of visual effects that result from various operators, including some that are not easily recreated with standard image editing tools. We find that this process allows for continuous, fine-grain control of image generation which can be helpful for creative applications. Next, we generate music-reactive videos using Stable Diffusion by passing audio features as parameters to network bending operators. Finally, we comment on certain transforms which radically shift the image and the possibilities of learning more about the latent space of Stable Diffusion based on these transforms.  
\end{abstract}

\section{Introduction}
\label{sec:intro}

We seek to create an artistic tool which aids in the creation of music visualizations: videos in which aspects of the image change in relation to aspects of the sound. We propose a system that generates music reactive videos given a sound file and some constraints. The system, which utilizes generative diffusion models \cite{Yang:24}, is flexible enough to create a wide variety of visual aesthetics. It can produce abstract textures and shapes as well as specific objects and scenes and can move between different visual aesthetics within the same video. Our hope is that the system creates a relationship between sound and image that is clear but complex. In this paper, we present preliminary steps towards these goals and investigate an implementation that shows promise while acknowledging that there is still more work to be done to create such a system.

Today, more and more artists work across disciplines and modalities, bridging the gaps between different types of media \cite{Condee:16, Tanya:17}. Various areas of study and artistic domains have sprung up at these intersections, such as audio-visual art \cite{Krupskyy:21, Edmonds:04}. From the perspective of a composer or musician, it may be desirable to bring other art forms, such as visual art, into one's practice \cite{Garro:12, Watkins:18}. Music visualizations can complement a piece of music by bringing it into a new modality.

One avenue for a composer to realize a music visualization is by collaborating with a visual artist. For example, the composer and artist Max Cooper, who is known for his music videos, works with a different visual artist for each of his videos. While these collaborations can be extremely fruitful and fulfilling for both sides, there can also be a desire for a single person to create both the sound and the visuals. This may lead to a more unified approach where the artist, working alone, can more fully realize their creative idea. 

In our opinion, it is important to note that by shutting themselves off from collaboration, the lone artist will be passing up opportunities to have their view of the piece expanded by working with another artist. It should not be forgotten that collaboration can be an extremely beneficial working method.

Nonetheless, if there is a desire to have more control over the creation of both the audio and visuals, then a composer may find themselves lacking the technical skills to create visual art; it is difficult for a single person to have expertise in both fields. Of course, this is not impossible as some individuals, such as artist Ryoji Ikeda, possess the skills to create both music and visual art. However by allowing one to create both visuals and music, it is possible for the artist to have conceptual unity across the two modalities. Nothing is lost in translation. 

In our proposed system, the artist can seek a specific visual aesthetic which is represented semantically using text or images. This aesthetic can be applied to the system as constraints on the generation of images. In order to apply these semantic constraints, we look to machine learning methods.

In the field of Music Information Retrieval (MIR), there has been a shift from using hand-crafted features to using machine-learned features, which has opened up new possibilities in audio representations \cite{Humphrey:13}. In the same way, we seek to push music visualization past the phase of hand-crafted one-to-one mappings, and into the area of machine learned mappings and semantics. Working in the pixel domain, just like working in the waveform domain, only allows certain operations or effects to be applied to an image or a sound. Just as one cannot remove the sound of one source from a complex auditory scene using standard audio methods, one cannot change the background of an image using standard image editing methods. In order to achieve such results, different methods are necessary; we need to work not at the pixel level but at the semantic level. It is this idea that guides our work.

As we will outline in Section \ref{sec:sota}, a standard approach to generating music visualizations is to map the audio features of the music to the visual features of the video. In this way, any number of audio features such as amplitude, pitch, noisiness, etc. can control any visual parameter such as color, brightness, etc. While this approach is a valid one, we search for a deeper mapping that is not a one-to-one, one-to-many, or many-to-one mapping from audio features to visual features. Our hope is that the complexity of this mapping allows for more meaningful and compelling visualizations.

In Section \ref{sec:methodology}, we begin this work by utilizing diffusion models for image generation. For a standard text-to-image diffusion model, the only control that a user has over the resulting image is the text prompt and the random seed. The random seed gives no expressive control to the user since there is no continuity; changing the seed by 1 has the same effect as changing it by 1,000. Therefore, the user can only control image generation through the text prompt. While this gives great semantic control, it does not give continuous or fine-grain control of the image. Small changes in the text prompt can lead to large changes in the resulting image. We seek a level of continuous control that follows the Lipschitz continuity: a small change in the input should lead to a small change in the output. For example, a visual effect such as saturation can be applied to an image, with a parameter giving fine-grain, continuous control. Since sounds are continuous and often have smooth changes, this is an important aspect of our system.

In order to reach this goal, we propose the application of network bending to be applied to pre-trained diffusion models as a method of exerting creative control over the output of the model. Network bending, proposed by \cite{Broad:22}, allows this control by applying transformations within the layers of the network during generation, giving the user the ability to influence output through one or multiple changing parameters.

As a first step, we investigate using network bending to affect the generation of images from a text-to-image diffusion model. We identify various functions that are capable of applying different visual transformations to images, illustrating that network bending can be applied to diffusion models. In Section \ref{ssec:experiments}, we list the different operators we experiment with and identify the visual effects that result from these operators in Section \ref{sec:discussion}.

In Section \ref{ssec:audiotovideo}, we seek to use network bending to generate an audio reactive video. This is done through frame by frame generation by an image generation model, where the creation of each frame is influenced by the current audio at the time the frame is displayed. The generated frames are then stitched together and the audio that conditioned the generation plays simultaneously. We generate short videos that take an audio file and text prompt as input. After choosing an operator to be used for network bending and an audio feature to be passed as a parameter to the operator, we create music-reactive videos. In the future, the operator and audio feature would not be hand-picked but machine-crafted, as we detail in Section \ref{sec:conclusion}.

To summarize, our contributions are as follows:
\begin{itemize}
    \item We show for the first time that network bending can be applied to diffusion models in order to exert expressive control over image generation
    \item We show the variety of visual effects that different transformations have on image output
    \item We show that videos can effectively be created using image generation models and that music reactive visualizations can be created using network bending
\end{itemize}

We provide our code at \url{https://github.com/dzluke/DAFX2024}. A series of videos and supplementary images that we generated can be viewed at \url{https://dzluke.github.io/DAFX2024/}.

\section{State of the Art}
\label{sec:sota}

A music visualization is the realization of a sonic and time-based phenomenon through light, color, shapes, or symbols. There are many different approaches to music visualization: the use of video and animation, lights and lasers, created using software or hardware. Visualizations can be static images or dynamic videos. They can be created in real-time or pre-computed, composed or algorithmically generated. In this paper, we will focus on visualizations which are digital and created using software. We will identify systems that are generative and can be real-time or offline.

Broadly speaking, visualizations fall into two categories: functional and aesthetic \cite{Panda:21}. Common in MIR, the goal of a functional visualization is to provide new information to the viewer, aid in analysis of a sound, or show the sound in a new light \cite{Lima:21}. Aesthetic visualization, on the other hand, is concerned with the creation of art. In this paper, we seek the aesthetic visualization of sound; our goal is to create art.

The line between functional and aesthetic visualizations can be blurred. For example, the spectrogram itself is a type of music visualization; often thought of as functional but used for artistic means as well \cite{Monacchi:13}. Martin Wattenberg's "The Shape of Song" toes the line between aesthetic and functional, visualizing the form of different musical pieces by connecting repeated sections in a way that reveals something new about the piece in an artistic way\footnote{\url{https://www.turbulence.org/Works/song/}}.

\subsection{Classical Methods}

Many methods have been employed to create both functional and aesthetic visualizations. Often they involve an analysis of a sound and a representation of that analysis through visual forms. For example, in \cite{Foote:99} similarities between different sections in a piece of music can be visualized by calculating the MFCCs of a segment and computing a similarity measure to all other segments in the piece. The self-similarity matrix that arises out of this is visualized as an image. In \cite{Cooper:06} the authors apply PCA to audio features and then use self-similarity and self-organizing maps to achieve various visualization methods, some real-time, for the purposes of music classification.

Within the realm of aesthetic visualization, a common approach to creating dynamic music visualizations is for the artist to create a mapping from audio features to visual features \cite{Bain:08, Vieira:12}. For example, the amplitude and spectral centroid could be mapped to the color and texture, respectively, of some objects on screen. The BPM could control a rate of movement of these objects, and, using beat detection, they could move around the screen on the beat. This mapping could be saved as a preset and different presents could be used for different types of music. This approach to visualization can be used for both real-time and pre-computed visualizations.

Common softwares for creating real-time visualizations include Jitter and TouchDesigner, which allow the creation and linking of modules that perform different computational tasks and are fast enough to create images on the fly \cite{Manzo:11}. Many libraries and plug-ins, such as Vsynth\footnote{\url{https://www.kevinkripper.com/vsynth}} for MaxMSP and Scintillator\footnote{\url{https://scintillatorsynth.org/}} for SuperCollider, allow artists to create visuals through preset or custom functions and can take input from any number of audio streams or sensor-based sources.

\subsection{Learning-based Methods}

Another approach to creating visualizations is through the use of machine learning, which can be used to generate images and videos. Generative Adversarial Networks (GANs), which consist of a discriminator network and a generator network, are able to generate images of a single class \cite{Goodfellow:14} and have been employed in various ways to create music visualizations. In \cite{Lee:20}, the authors train an encoder-decoder model to perform music-to-image, and then use the resulting image to apply style to an input image in a style-transfer process. A major limitation of this approach is that the stylization effect does not change over time, but determines a single visual style based on the musical input, and is therefore not suited for dynamic music visualization. Another GAN-based approach, Tr\"aumerAI \cite{Jeong:21}, uses a CNN as a music encoder and translates music embeddings into a visual embedding space, and then generates images using StyleGAN2 \cite{Karras:19}. This approach is effective for creating dynamic visualizations, however there does not appear to be a strong temporal synchronization between audio and video. In a system that resembles our goal, the authors of \cite{Brouwer:20} apply network bending and other techniques to StyleGAN2 to create music reactive videos that change based on various audio features. The major difference between this system and the one we propose is that we use diffusion models and hope to create a system that does not use hand-picked mappings between audio features and visual characteristics.

While some systems show promise for generalized visualization, we avoid using GANs for a number of reasons. First, GANs are usually specific to one style or object type, like paintings, puppies, or Van Gogh, instead of a generalized image feature space. The reason for this is that we wish the user to be able to move between visual aesthetics within one video; for example from an impressionist painting to a 3D rendering to a graphite drawing. Standard GANs are also unable to generate images conditioned on text or image, which gives the user less semantic control over the generation.

More recently, diffusion models have been employed for image generation. Diffusion models work by training a network to remove noise from images, and when pure noise is fed to the model it can be guided by a text prompt to generate an image of that prompt \cite{Ho:20}. These models have been used for creating music visualizations in a number of ways. AudioToken \cite{Yariv:23} is capable of performing audio-to-image, generating an image that reflects the source of a sound, such as a picture of a bird when a bird song is input. MM-Diffusion \cite{Ruan:23} jointly generates video and audio, for example generating a video of the ocean and the sounds of waves lapping at the shore. However, both of these examples are functional visualizations, aiming to create an image that provides information about a sound, but we seek aesthetic visualizations.
 
Generative Disco \cite{Liu:23} is the closest example to our goal: a video that moves through different text prompts is generated, and the interpolation speed between text prompts is determined by the amplitude of percussive elements at a given point in time. This system is built from a modified version of Stable Diffusion capable of generating music reactive videos\footnote{\url{https://github.com/nateraw/stable-diffusion-videos}}. The main drawbacks are that the only visual feature that is changing is the interpolation between prompts and the only audio feature being employed is the amplitude. We seek a system that has a complex relationship between the timbral elements of the audio and the visual characteristics of the image, not a one-to-one mapping.

\section{Methodology}
\label{sec:methodology}
\begin{figure}[ht!]
    \centering
    \includegraphics[width=\columnwidth]{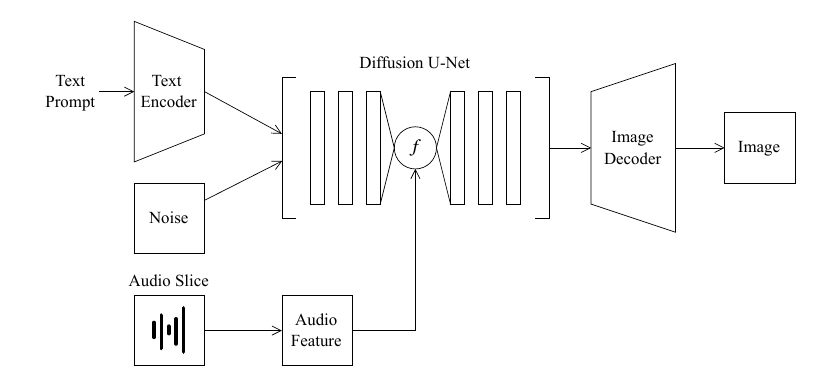}
    \caption{A block diagram of our system. An operator $f$ is inserted at some layer in the U-Net. The audio feature computed from the audio at that time is passed as the parameter to $f$.}
    \label{fig:block-diagram}
\end{figure}
In the previous section, we saw how diffusion models have shown results for image generation and promise for audio visualization. Therefore, we use Stable Diffusion, an open-source text-to-image diffusion model, to generate all examples shown in this paper \cite{Rombach:22}. Stable Diffusion can generate images in multiple ways; the methods relevant to us are text-to-image and image-to-image. The architecture of Stable Diffusion consists of three distinct networks: a text encoder, a diffusing U-Net, and an image decoder.

Network Bending is applied in the layers of the U-Net, which is where the diffusion process takes place. At any point in the diffusion process the image being diffused is represented by a compressed encoding which is a tensor of shape (4, 64, 64). These tensors are then input to an operator and the transformed output is fed to the next layer. By applying network bending, we enable parameterized control of the output image, which is not possible otherwise.

There are four parameters that define an individual application of network bending:

\begin{enumerate}
    \item Layer: the operator can be applied before or after any layer of the network
    \item Operator: can be a point-wise, tensor-wise, or morphological transformation
    \item Parameter: most operators take a parameter as input, such as the scalar to multiply by or the angle to rotate by
    \item Feature: the operator can be applied to all elements of the latent tensor, a single dimension of the tensor, or a random selection of its features
\end{enumerate}
Each of these parameters can have an effect on the resulting image \cite{Broad:22}.

\subsection{Experiments}
\label{ssec:experiments}

In order to test the different parameters that can affect network bending, we first generate images. Each image has one transform applied at one layer. Many of the transformations we apply are taken from \cite{Broad:22}. We perform a grid search on the parameter space of each operator, and disregard parameters which lead to images that are entirely black. Unless otherwise noted, the point-wise functions are applied to every element of the latent tensor. 

All images are generated using pre-trained Stable Diffusion v1\footnote{\url{https://github.com/CompVis/stable-diffusion}} with the frozen v1.4 checkpoint\footnote{\url{https://huggingface.co/CompVis/stable-diffusion-v-1-4-original}}; we do not perform any additional training or fine-tuning. We use the DDIM sampler with the default setting of 50 sampling steps and the seed set to 46. In our experiments it takes approximately 1 second to generate a single frame on an NVIDIA GeForce RTX 4090, meaning a one minute video at 20 FPS takes approximately 20 minutes to generate. 

\subsubsection{Point-wise Operations}

We apply numerous point-wise operators which transform each element of the latent tensor and select four operators that lead to meaningful visual change in the resulting image. For each function, the input $x$ is one element of the latent tensor, and $r$ is a parameter of the given operator.

\begin{enumerate}
    \item Addition of a scalar: \(f(x) = x + r \)
    \item Multiplication by a scalar: \(f(x) = x \cdot r\)
    \item Hard threshold: \(f(x) = \begin{cases}
                        1 & \text{if } x \geq r \\
                        0 & \text{otherwise}
                    \end{cases}\)
    \item Inversion: \(f(x) = \frac{1}{r} - x\)
\end{enumerate}

\def\imagescale{0.15}

\begin{figure*}[ht!]
     \centering
     \begin{subfigure}[t]{0.19\textwidth}
         \centering
         \includegraphics[scale=\imagescale]{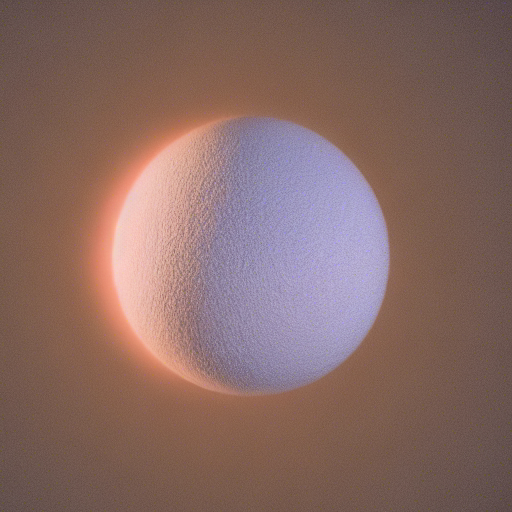}
         \caption{No operator applied}
         \label{fig:pointwise-a}
     \end{subfigure}
     \hfill
     \begin{subfigure}[t]{0.19\textwidth}
         \centering
         \includegraphics[scale=\imagescale]{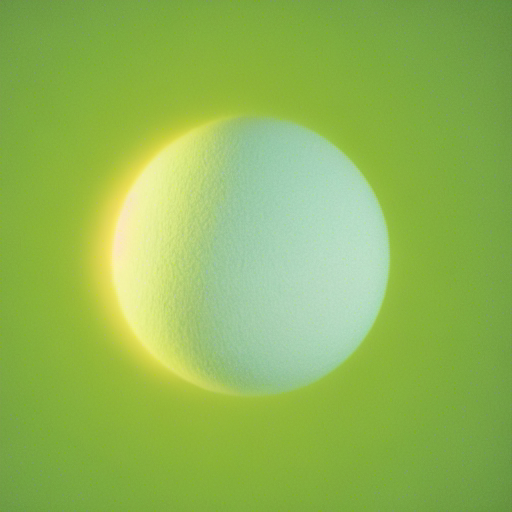}
         \caption{Add scalar, $r=1$, layer 40}
         \label{fig:pointwise-b}
     \end{subfigure}
     \hfill
     \begin{subfigure}[t]{0.19\textwidth}
         \centering
         \includegraphics[scale=\imagescale]{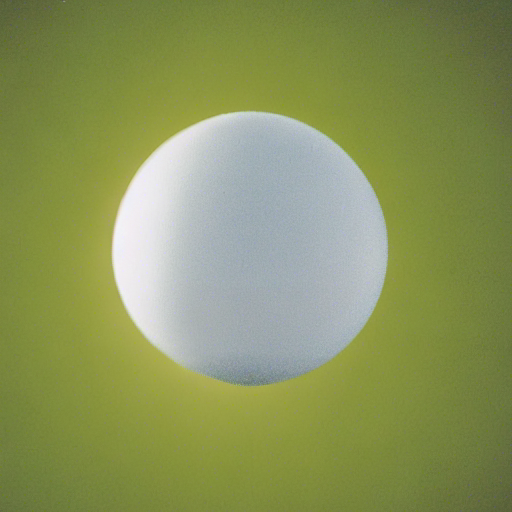}
         \caption{Add scalar, $r=1$, layer 0, random selection of 5\% of the features}
         \label{fig:pointwise-c}
     \end{subfigure}
     \hfill
     \begin{subfigure}[t]{0.19\textwidth}
         \centering
         \includegraphics[scale=\imagescale]{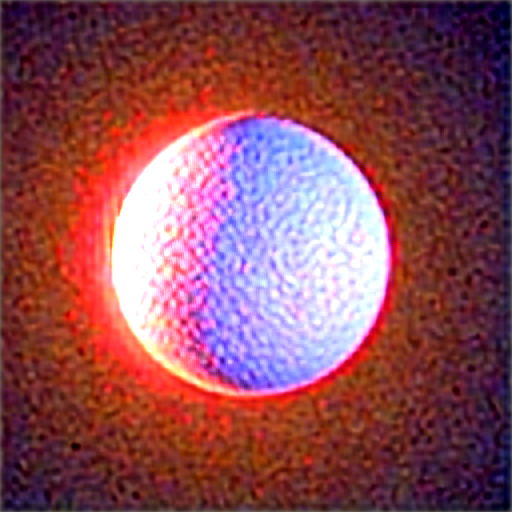}
         \caption{Multiply scalar, $r=10$, layer 49}
         \label{fig:pointwise-d}
     \end{subfigure}
     \hfill
     \begin{subfigure}[t]{0.19\textwidth}
         \centering
         \includegraphics[scale=\imagescale]{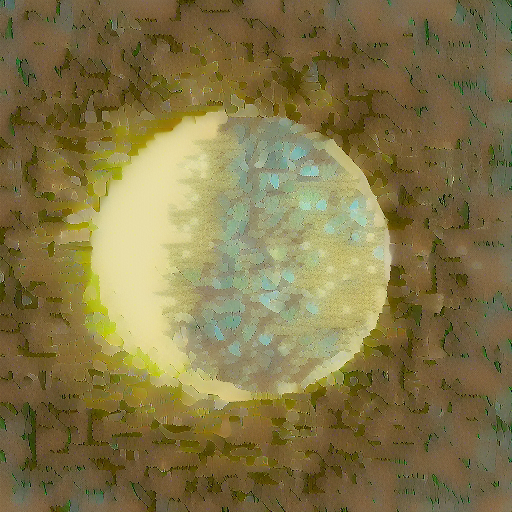}
         \caption{Hard threshold, $r=0.1$, layer 49}
         \label{fig:pointwise-e}
     \end{subfigure}
     \begin{subfigure}[t]{0.19\textwidth}
         \centering
         \includegraphics[scale=\imagescale]{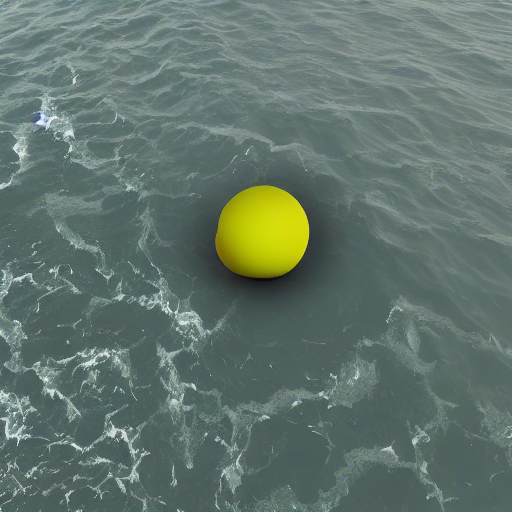}
         \caption{Inversion, $r=1000$, layer 0}
         \label{fig:pointwise-f}
     \end{subfigure}
     \hfill
     \begin{subfigure}[t]{0.19\textwidth}
         \centering
         \includegraphics[scale=\imagescale]{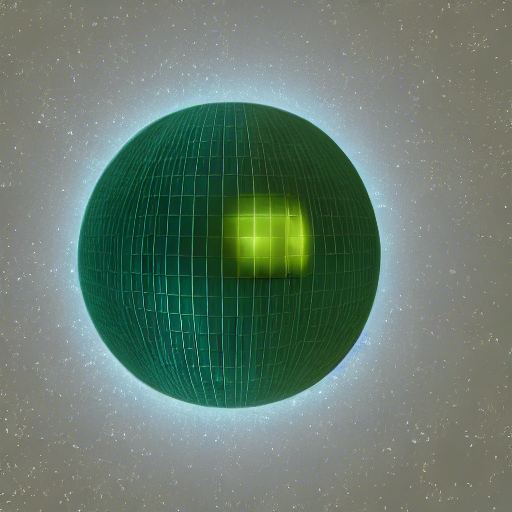}
         \caption{Inversion, $r=1000$, layer 3}
         \label{fig:pointwise-g}
     \end{subfigure}
     \hfill
     \begin{subfigure}[t]{0.19\textwidth}
         \centering
         \includegraphics[scale=\imagescale]{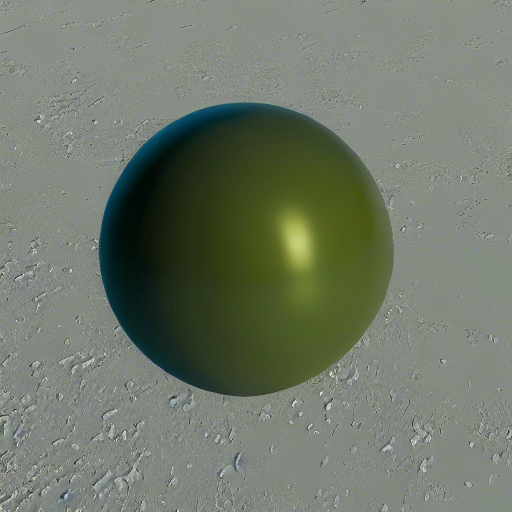}
         \caption{Inversion, $r=1000$, layer 6}
         \label{fig:pointwise-h}
     \end{subfigure}
     \hfill
     \begin{subfigure}[t]{0.19\textwidth}
         \centering
         \includegraphics[scale=\imagescale]{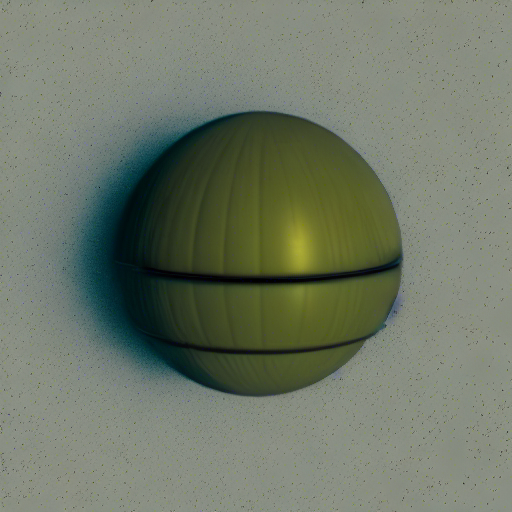}
         \caption{Inversion, $r=1000$, layer 9}
         \label{fig:pointwise-i}
     \end{subfigure}
     \hfill
     \begin{subfigure}[t]{0.19\textwidth}
         \centering
         \includegraphics[scale=\imagescale]{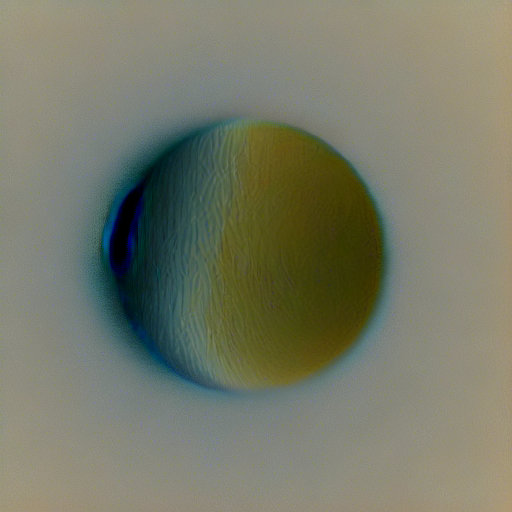}
         \caption{Inversion, $r=1000$, layer 49}
         \label{fig:pointwise-j}
     \end{subfigure}
        \caption{Image generations using the prompt "a floating orb" with various point-wise operators applied}
        \label{fig:pointwise}
\end{figure*}

\subsubsection{Tensor Operations}

Another type of transformation we experiment with are tensor operations, in which an operator tensor is contracted with the latent tensor, applying an operation in the same way as a matrix multiplication would. These operations can be thought of as shifting the latent tensor to a new location in the feature space. The two tensor operations we apply are rotation and reflection.

Rotation is applied by contracting a rotation matrix with the latent tensor. We experiment with the following 4x4 matrices:
\def\rotatex{R_1}
\def\rotatey{R_2}
\def\rotateytwo{R_3}
\def\rotatez{R_4}
\renewcommand{\arraystretch}{0.6}
\[
\rotatex = 
\begin{bmatrix}
    1 & 0 & 0 & 0 \\
    0 & \cos{\theta} & -\sin{\theta} & 0 \\
    0 & \sin{\theta} & \cos{\theta} & 0 \\
    0 & 0 & 0 & 1
\end{bmatrix}
%
\rotatey = 
\begin{bmatrix}
    \cos{\theta} & 0 & \sin{\theta} & 0 \\
    0 & 1 & 0 & 0 \\
    -\sin{\theta} & 0 & \cos{\theta} & 0 \\
    0 & 0 & 0 & 1
\end{bmatrix}
\]
\[
\rotateytwo = 
\begin{bmatrix}
    \cos{\theta} & 0 & 0 & \sin{\theta} \\
    0 & 1 & 0 & 0 \\
    0 & 0 & 1 & 0 \\
    -\sin{\theta} & 0 & 0 & \cos{\theta} 
\end{bmatrix}
%
\rotatez =
\begin{bmatrix}
    \cos{\theta} & -\sin{\theta} & 0 & 0 \\
    \sin{\theta} & \cos{\theta} & 0 & 0 \\
    0 & 0 & 1 & 0 \\
    0 & 0 & 0 & 1
\end{bmatrix}
\]

Reflection is applied through four different 4x4 reflection matrices, in which each one reflects across one dimension. A reflection matrix is the 4x4 identity matrix with one of the elements on the diagonal set to $-1$.

\begin{figure*}[ht!]
     \centering
     \begin{subfigure}[t]{0.245\textwidth}
         \centering
         \includegraphics[scale=\imagescale]{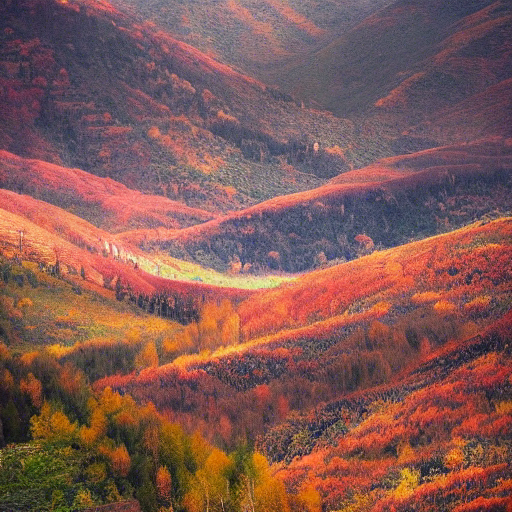}
         \caption{No operator applied}
         \label{fig:rotate-a}
     \end{subfigure}
     \hfill
     \begin{subfigure}[t]{0.245\textwidth}
         \centering
         \includegraphics[scale=\imagescale]{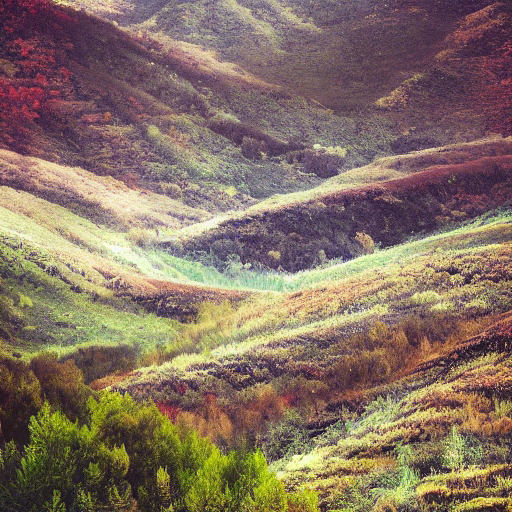}
         \caption{$\theta = 0.25\pi$}
         \label{fig:rotate-b}
     \end{subfigure}
     \hfill
     \begin{subfigure}[t]{0.245\textwidth}
         \centering
         \includegraphics[scale=\imagescale]{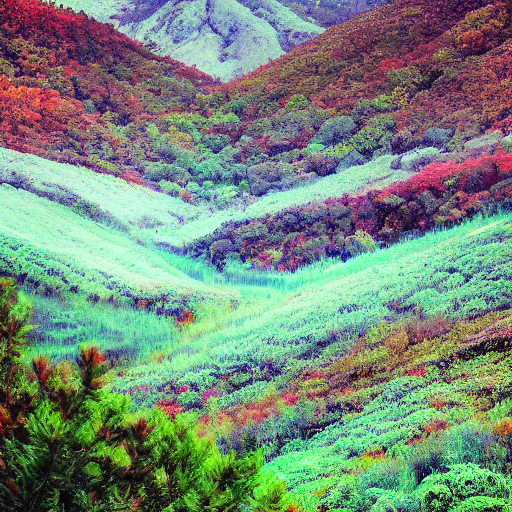}
         \caption{$\theta = 0.5\pi$}
         \label{fig:rotate-c}
     \end{subfigure}
     \hfill
     \begin{subfigure}[t]{0.245\textwidth}
         \centering
         \includegraphics[scale=\imagescale]{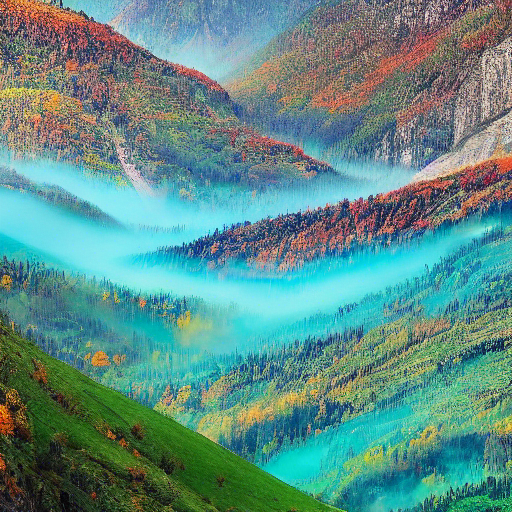}
         \caption{$\theta = 0.75\pi$}
         \label{fig:rotate-d}
     \end{subfigure}
     \hfill
     \begin{subfigure}[t]{0.245\textwidth}
         \centering
         \includegraphics[scale=\imagescale]{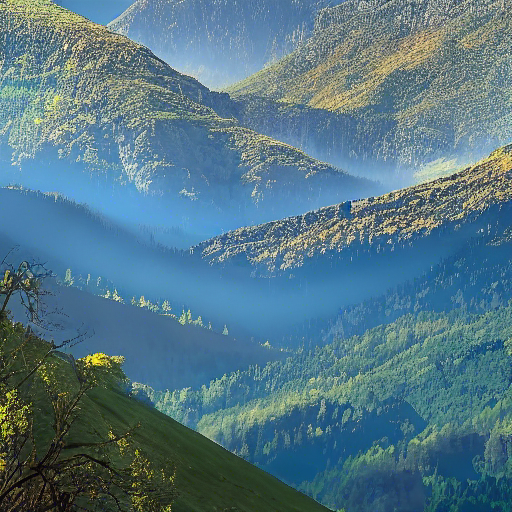}
         \caption{$\theta = \pi$}
         \label{fig:rotate-e}
     \end{subfigure}
     \begin{subfigure}[t]{0.245\textwidth}
         \centering
         \includegraphics[scale=\imagescale]{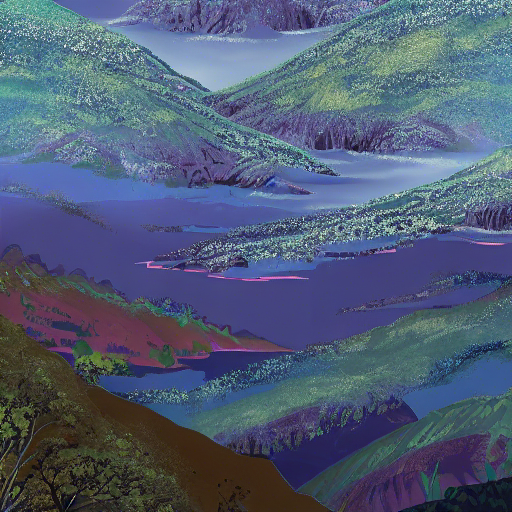}
         \caption{$\theta = 1.25\pi$}
         \label{fig:rotate-f}
     \end{subfigure}
     \hfill
     \begin{subfigure}[t]{0.245\textwidth}
         \centering
         \includegraphics[scale=\imagescale]{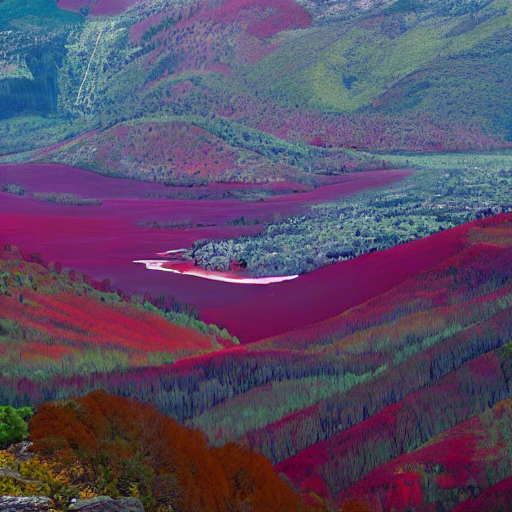}
         \caption{$\theta = 1.5\pi$}
         \label{fig:rotate-g}
     \end{subfigure}
     \hfill
     \begin{subfigure}[t]{0.245\textwidth}
         \centering
         \includegraphics[scale=\imagescale]{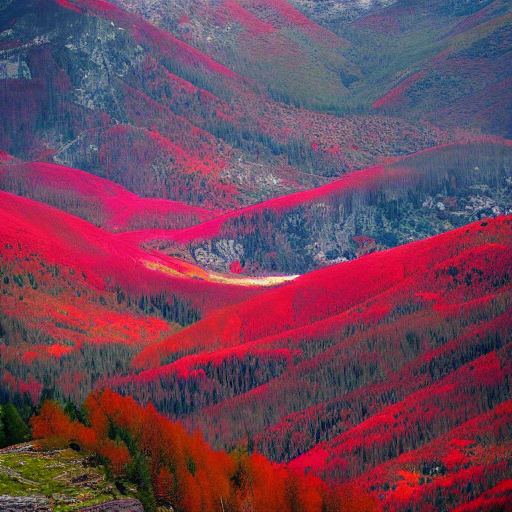}
         \caption{$\theta = 1.75\pi$}
         \label{fig:rotate-h}
     \end{subfigure}
        \caption{Image generations using the prompt "a gorgeous landscape" with $\rotatex$ applied at layer 20 with changing angle}
        \label{fig:rotate}
\end{figure*}

\begin{figure*}[ht!]
     \centering
     \begin{subfigure}[t]{0.24\textwidth}
         \centering
         \includegraphics[scale=\imagescale]{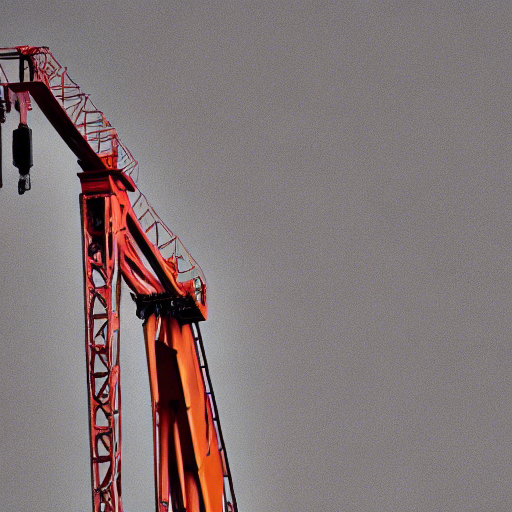}
         \caption{"a photo of a crane"}
         \label{fig:y equals x}
     \end{subfigure}
     \hfill
     \begin{subfigure}[t]{0.24\textwidth}
         \centering
         \includegraphics[scale=\imagescale]{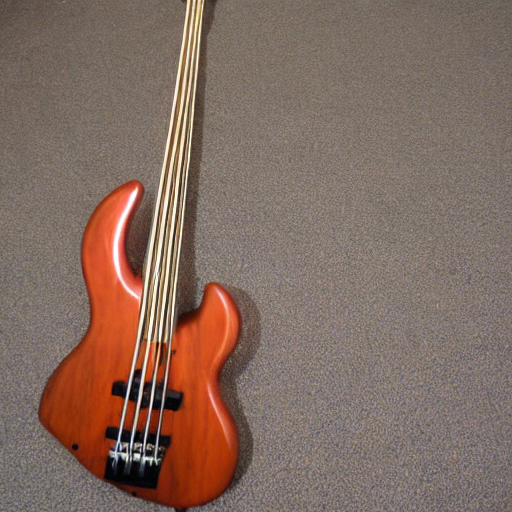}
         \caption{"bass"}
         \label{fig:y equals x}
     \end{subfigure}
     \hfill
     \begin{subfigure}[t]{0.24\textwidth}
         \centering
         \includegraphics[scale=\imagescale]{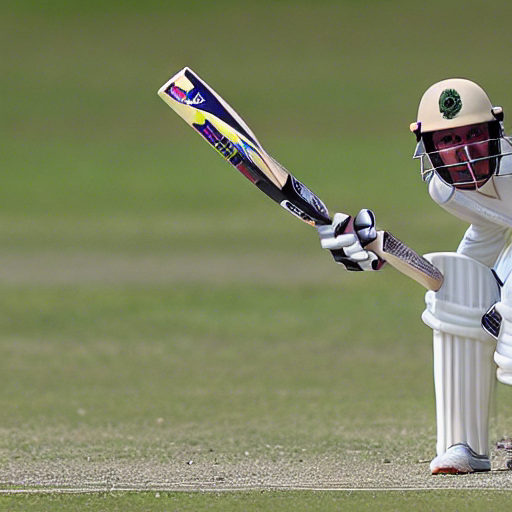}
         \caption{"cricket"}
         \label{temp}
     \end{subfigure}
     \hfill
     \begin{subfigure}[t]{0.24\textwidth}
         \centering
         \includegraphics[scale=\imagescale]{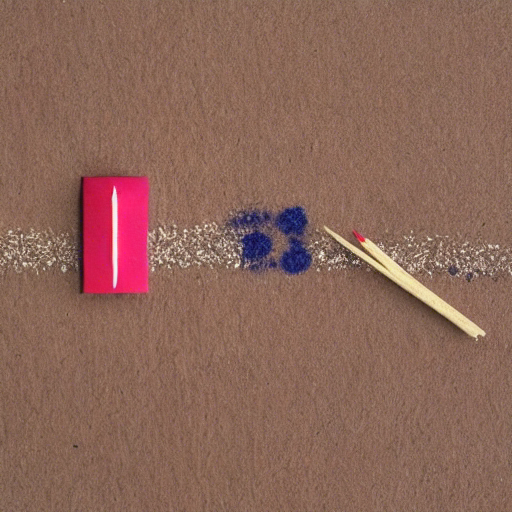}
         \caption{"a match"}
         \label{fig:five over x}
     \end{subfigure}
     \hfill
     \begin{subfigure}[t]{0.24\textwidth}
         \centering
         \includegraphics[scale=\imagescale]{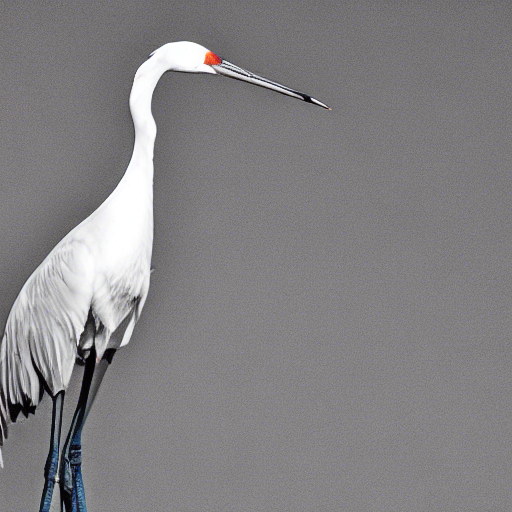}
         \caption{$\theta = 0.5\pi$}
         \label{fig:five over x}
     \end{subfigure}
     \hfill
     \begin{subfigure}[t]{0.24\textwidth}
         \centering
         \includegraphics[scale=\imagescale]{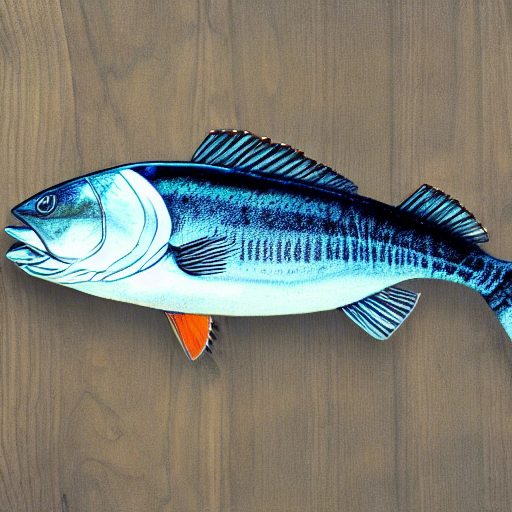}
         \caption{$\theta = 0.75\pi$}
         \label{fig:five over x}
     \end{subfigure}
     \hfill
     \begin{subfigure}[t]{0.24\textwidth}
         \centering
         \includegraphics[scale=\imagescale]{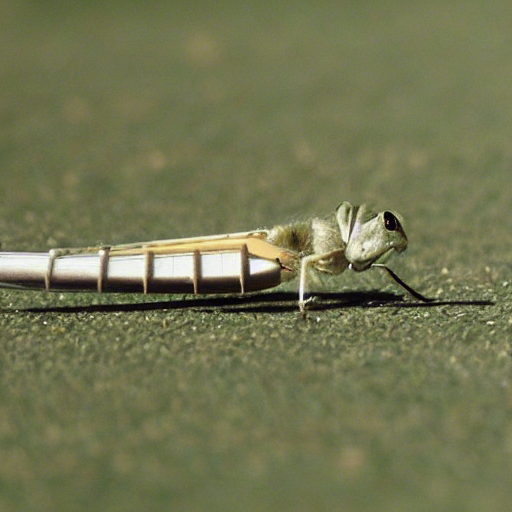}
         \caption{$\theta = 0.75\pi$}
         \label{fig:five over x}
     \end{subfigure}
     \hfill
     \begin{subfigure}[t]{0.24\textwidth}
         \centering
         \includegraphics[scale=\imagescale]{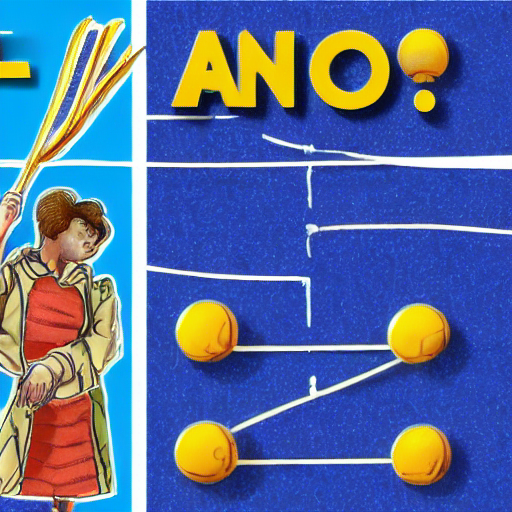}
         \caption{$\theta = 1.25\pi$}
         \label{fig:five over x}
     \end{subfigure}
        \caption{Examples of semantic shift. The top image is generated with no transformation applied. The bottom image is the same prompt but with $\rotatex$ applied at layer 0}
        \label{fig:semantic shift}
\end{figure*}

\subsubsection{Morphological Transformations}

Finally, we employ two morphological transforms, erosion and dilation, as found in \cite{Broad:22}. These are applied to the latent tensor, treating it as a 4-channel image. The transformations are implemented using the Kornia library \cite{Riba:20}. 

Overall these transformations did not lead to as meaningful results as achieved in \cite{Broad:22}, however we found that normalizing the tensor after applying the transformation led to more promising results. The normalization is done by subtracting the mean from each element, but applied to specific dimensions. For example, if dimension 1 is normalized, then the mean of each row is subtracted from each element in that row.

\subsection{Audio-to-Video}
\label{ssec:audiotovideo}

After investigating the visual effects that result from different transformations, we use Stable Diffusion to generate videos in two distinct ways: using text-to-image with batched noise and using image-to-image with the previous frame as input. 

The first method for creating videos uses standard text-to-image generation but changes the initial noise that is input to the system \cite{Stenbit:22}. The initial noise is generated in the following way: first a standard normal distribution is sampled to create a two tensors of noise, which we call $A$ and $B$.
Then, to generate frame $i$ out of total of $k$ frames, the initial noise passed to the model equals $A * \sin{\frac{2\pi i}{k}} + B * \cos{\frac{2\pi i}{k}}$. When generating videos with text-to-image, the user supplies a single text prompt or two text prompts to interpolate between. If two prompts are given, the video will start at the first prompt and end at the second prompt. This image interpolation is achieved through linearly interpolating between the text encodings of the two prompts.

The second method uses image-to-image to create videos. Image-to-image is a process in which the input to the diffusion U-Net is a text prompt, an initialization image, and a "strength" parameter. The diffusion process starts from the initialization image, which has had noise added to it \cite{Rombach:22}. The amount of noise added is in relation to the strength parameter: a value of 0 corresponds to no noise being added to the image and a value of 1 means the initialization image will be turned into complete noise and have no effect on the resulting image.

To generate videos using image-to-image, the user must provide either an initialization image or a text prompt. The first frame of the video is either the initialization image or the image generated from passing the text prompt to text-to-image. The generation of each subsequent frame is conditioned on the previous frame, using image-to-image, and with an empty string as the text prompt. 

For both methods, we achieve audio-reactivity in the video by applying network bending during the generation of each frame with a user defined operator and audio feature. For a given frame, there is a 50 millisecond window of audio that will play while that frame is shown. The chosen audio feature is calculated for this specific window of audio and is then passed as a parameter to the chosen operator. Usually, the value must be scaled to a different range, as the range of values that give meaningful results for a given operator is not necessarily the same range of the audio feature. We experiment with different audio features including RMS, spectral shape (centroid, spread, skewness, kurtosis), and spectral flux. We choose these features because they are commonly used in MIR tasks and can represent audio with a single value, which is useful since our transformations take only one parameter \cite{Peeters:11}.

For example, we generate a video\footnote{The video can be viewed at \url{https://dzluke.github.io/DAFX2024/}} using text-to-image with solo piano as audio input, the prompt "3D mesh geometry," and apply the rotation $\rotatex$ at layer 40 with the RMS of the audio being passed as the angle of rotation. The RMS is scaled to the range $0$ to $2\pi$ before being passed to the operator. On each note onset from the piano, the image responds through shifting colors which emerge from the black and white mesh. These visual changes follow the amplitude envelope of the piano, with a strong shift in color at the attack and a decay to the original image with the piano. When the piano is quiet or silent, the black and white mesh continues to change as a result of the initialization noise that is fed to the generation.


\section{Discussion}
\label{sec:discussion}

In order to create music-reactive videos, we need to identify the visual effect that each transformation has on the image. We find that various operators are capable of a number of different effects. A green color filter can be achieved through adding a scalar (Figure \ref{fig:pointwise-b}), and a saturation effect is achieved through multiplication by a scalar (Figure \ref{fig:pointwise-d}). These are standard visual effects that are achievable with media editing software. However, other transforms lead to results that are more complex and not accessible through standard methods. For example, adding a scalar to only 5\% of the features can change only the background color of the image (Figure \ref{fig:pointwise-c}) and applying a hard threshold before the last layer creates a stained glass effect (Figure \ref{fig:pointwise-e}). 

\begin{figure*}[ht!]
     \centering
     \begin{subfigure}[t]{0.329\textwidth}
         \centering
         \includegraphics[scale=\imagescale]{images/a_floating_orb.png}
         \label{fig:y equals x}
     \end{subfigure}
     \hfill
     \begin{subfigure}[t]{0.329\textwidth}
         \centering
         \includegraphics[scale=\imagescale]{images/a_gorgeous_landscape.png}
         \label{fig:y equals x}
     \end{subfigure}
     \hfill
     \begin{subfigure}[t]{0.329\textwidth}
         \centering
         \includegraphics[scale=\imagescale]{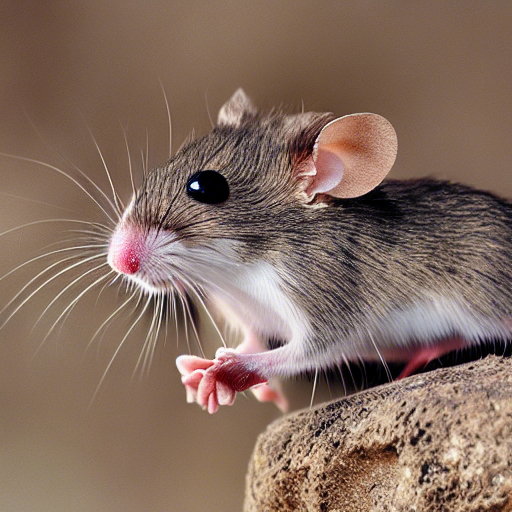}
         \label{fig:three sin x}
     \end{subfigure}
     \hfill
     \begin{subfigure}[t]{0.329\textwidth}
         \centering
         \includegraphics[scale=\imagescale]{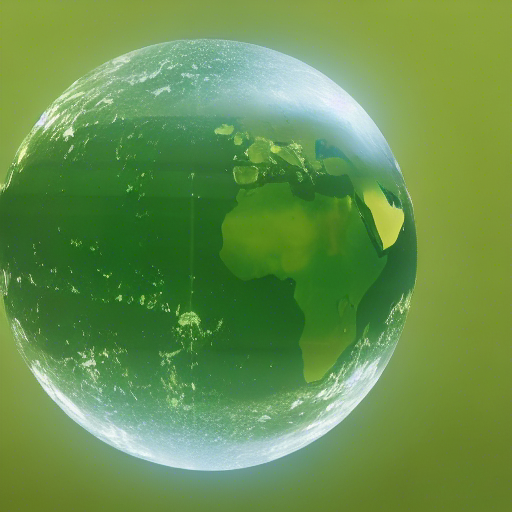}
         \caption{prompt: "a floating orb" \\ inversion, $r=30$, layer 0}
         \label{fig:five over x}
     \end{subfigure}
     \hfill
     \begin{subfigure}[t]{0.329\textwidth}
         \centering
         \includegraphics[scale=\imagescale]{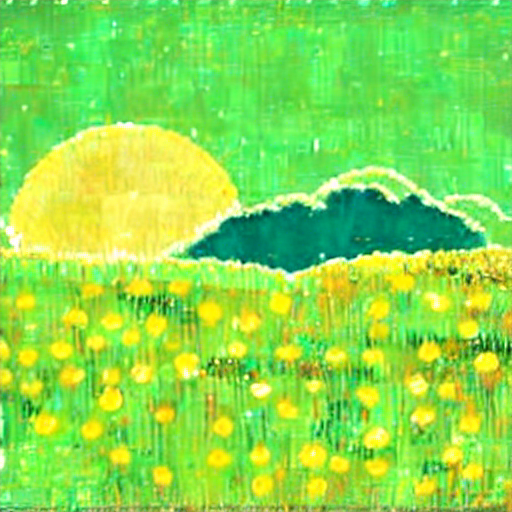}
         \caption{prompt: "a gorgeous landscape" \\ add scalar, $r=2$, layer 0, applied to 5\% of features}
         \label{fig:five over x}
     \end{subfigure}
     \hfill
     \begin{subfigure}[t]{0.329\textwidth}
         \centering
         \includegraphics[scale=\imagescale]{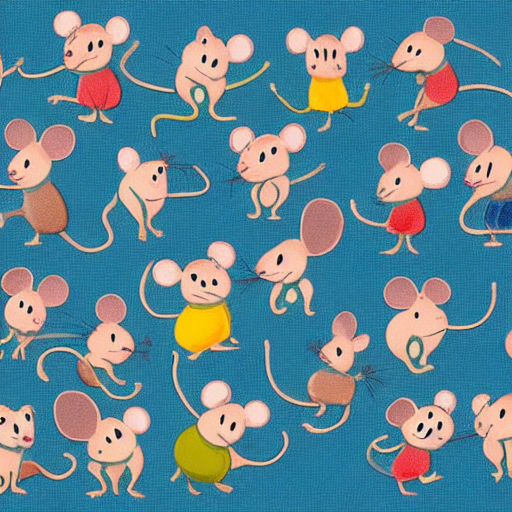}
         \caption{prompt: "mice" \\ rotation by $\rotatex$, $\theta=\pi$, layer 0}
         \label{fig:five over x}
     \end{subfigure}
        \caption{Examples of scene change with various prompts. The upper image has no transformation applied.}
        \label{fig:scene-change}
\end{figure*}

The result of applying inversion, as shown in Figures \ref{fig:pointwise-f}  - \ref{fig:pointwise-j}, leads to a shift in the image which is larger than a filter effect. We call this a "scene change": a transformation in which coherency is maintained but a significant shift in the image's contents or style has occurred. As we see in Figure \ref{fig:pointwise-f}, applying inversion before the first layer places the orb in a background of ocean water while changing the size, location, color, and texture of the orb. We find that multiple transformations are capable of creating scene changes, as can be seen in Figure \ref{fig:scene-change}.

As seen in Figure \ref{fig:rotate}, applying rotations as tensor-wise operations cycles the image through various color filters. The range of possible colors is determined by the rotation matrix and the color change is a result of the angle of rotation. Interestingly, as the angle increases from $0$ to $2\pi$ in $\rotatex$, the color filter moves along the color spectrum: yellow, green, blue, indigo, violet, red. While the effect may be a simple color filter, it sometimes does more than this: in Figure \ref{fig:rotate-f}, a blue filter is applied to the image and the creation of blue lakes appears in what were previously valleys. The other rotation matrices apply a similar color filter, but only between a few colors instead of the full spectrum. For example, $\rotatey$ applies an orange or blue filter, depending on the angle. $\rotateytwo$ applies a brown filter and $\rotatez$ applies a purple filter. Similar to a rotation, a reflection across one dimension applies a color filter: no change for the first dimension, purple for the second dimension, blue for the third dimension, and orange for the fourth dimension\footnote{See examples at \url{https://dzluke.github.io/DAFX2024/}}.

When a rotation is applied to a text prompt that is a homograph, words with the same spelling but different meanings, we find that the semantic meaning of the image may change. We call this effect a "semantic shift." In Figure \ref{fig:semantic shift}, this effect can be seen with different text prompts and various angles of rotation. When the prompt "a photo of a crane" is generated with no transformation applied, Stable Diffusion creates an image of a mechanical crane used in construction. If the same prompt is used but a rotation with $\rotatex, \theta=0.5\pi$ is applied, an image of a crane bird is generated. Similar results occur with the prompts "bass" and "cricket." For the prompt "a match," we have an image of a matchstick and an abstract image representing some imagined game: we see four balls and a person holding a bat. A semantic shift also occurs from the reflection and inversion operators. This is consistent because a reflection is similar to a rotation by $\pi$ and inversion is similar to a reflection across all dimensions.

The morphological operators of erosion and dilation lead to a kaleidoscope-like effect at early layers and a blurring effect at later layers when normalization is applied after the operator. When used with the prompt "a floating orb" at early layers, smaller, multi-colored duplicate spheres are created around the central orb.

We also experiment with applying transformations to only certain dimensions of the latent vector. When a scalar is added only to the middle row of the tensor, a green bar appears across the middle of the image. This suggests that there is a relationship between the spatial layout of the compressed tensor and the resulting image. Therefore, it may be possible to apply transformations to only a specific part of the image, while leaving the rest of the image untouched.

\section{Conclusions and Future Work}
\label{sec:conclusion}

We propose a tool for the creation of music visualization videos using deep learning. We find that network bending can successfully be applied to Diffusion Models and shows promise for allowing continuous, fine-grained control of image generation and the creation of videos. There is a wide range in the complexity of effects that different transformations lead to. Some transformations lead to simple effects such as color filtering or image saturation, yet we also achieve transformations that are considerably more advanced: scene changes and semantic shifts. These effects can be produced through different operators, including inversion, rotation, reflection, and adding a scalar. These advanced effects are a strong capability of our system since they are not easily achieved through standard image editing tools.

Through our experiments, we find some generalities on the effects of different transforms on the latent space. In general, increasing or decreasing the value of the latent tensor leads to changes in color. Increasing, through addition or thresholding, results in an image with more green in it, and decreasing results in the image becoming more purple. 

Applying transforms to earlier layers, especially before the first layer, leads to the most dramatic change in the resulting image. This is due to the fact that at later layers in the diffusion process, the image has been mostly formed. At early layers, there is still a potential for a significant shift in the image, since it is still predominantly noise. The scene changes and semantic shifts we see occur only if the transform is applied at the earliest layers. Applying certain transformations at the last layer can be useful to apply a specific visual effect while keeping the coherency of the original image.

While experimenting with different transformations, we find that some tensor operations can lead to a "semantic shift" when the text input is a homograph. This may suggest that concepts which are linked by the same word are laid out in the latent space in a relationship that can be accessed through geometric manipulations. The possibility of a geometry of information \cite{Cont:11} in the latent space of Stable Diffusion is extremely preliminary but is an interesting byproduct of our work and may be a path forward for gaining more understanding of the latent space of Stable Diffusion. The authors find it interesting that one of the most simple effects, color filter, and one of the most complex, semantic shift, are a result of the same operation.

As we noted earlier, the work shown in this paper are the first steps towards the realization of a system for music visualization. An important next step is to employ machine-crafted operators instead of hand-picked transforms. One possible approach to this is to feed the audio into an auto-encoder which outputs a compressed encoding, which is then applied as an operator on the latent tensor. We would also like the semantic constraints applied by the user to be a collection of text, images, or videos. These constraints could define a subspace of the latent space that is navigated during image generation. Similarly, the user could provide specific time points at which each prompt is displayed, and our system could interpolate between these prompts, allowing for temporal and narrative control of the video. Another possible avenue to improve the quality of our videos could be through image upscaling techniques and applying smoothing to the audio features \cite{Kim:24}.



Furthermore, we would like to investigate the semantic shift that results from certain transforms in order to better understand the latent space of Stable Diffusion. The invariances of an operator may define a topology defined by the orbit of the operator. This could allow one to create connections between disconnected images through the chaining of operators.

It is difficult to apply quantitative measurements to properly assess the artistic output of our system. However, we would like to explore this further through employing video distance metrics \cite{Unterthiner:19, Martinez:19} and also assess our system qualitatively through user studies.

Finally, there is potential for applying network bending to different types of generative networks, including other image networks, which may give different results based on differences in the latent space. Our methodology could also be applied to a video generation network or music generation network \cite{Evans:24}, which could allow the user to have fine-grain control of the music, perhaps changing timbral, pitch-based, or rhythmic aspects of the music in a continuous way. We believe this could be a powerful tool for creative control of text-to-music models.

\section{Acknowledgments}
Special thanks to Xiaojian Sun, Halie Sung, and Juan Lucas Umali.

\nocite{*}
\bibliographystyle{IEEEbib}
\bibliography{DAFx24_tmpl} 

\end{document}